\documentclass[12pt]{iopart}
\usepackage{iopams}

\begin{document}

\title[Phase-space descriptions of operators and the Wigner distribution II]{Phase-space descriptions of operators and the Wigner distribution in quantum mechanics II. \mbox{The finite} dimensional case}

\author{S Chaturvedi\dag, E Ercolessi\ddag\footnote{To
whom correspondence should be addressed (ercolessi@bo.infn.it)}, G Marmo\S, G Morandi$\|$, \mbox{N Mukunda}\P\ and R Simon$^+$}

\address{\dag School of Physics, University of Hyderabad, Hyderabad 500 046, India}

\address{\ddag Physics Department, University of Bologna, INFM-CNR and INFN. Via Irnerio 46, I-40126, Bologna, Italy}

\address{\S Dipartimento di Scienze Fisiche, Universit'di Napoli and INFN. Via Cinzia, I-80126 Napoli, Italy}

\address{$\|$ Physics Department, University of Bologna, INFM-CNR and INFN. V.le B.Pichat 6/2, I-40127, Bologna, Italy}

\address{\P Centre for High Energy Physics, Indian Institute of Science, Bangalore 560 012, India}

\address{$^+$ The Institute of Mathematical Sciences, C.I.T. Campus, Chennai 600113, India}

\begin{abstract}

A complete solution to the problem of setting up Wigner distribution for
$N$-level quantum systems is presented. The scheme makes use of some of the 
ideas introduced by Dirac in the course of defining functions of noncommuting
observables and works uniformly for all $N$. Further, the construction
developed here has the virtue of being essentially input-free in that it
merely requires finding a square root of a certain $N^2\times N^2$
complex symmetric matrix, a task which, as is shown, can always be accomplished analytically. As an
illustration, the case of a single qubit  is considered in some detail 
and it is shown that one recovers the result of Feynman and Wootters 
for this case without recourse to any auxiliary constructs. 
\end{abstract}



\maketitle

\section{Introduction}

There has been considerable interest for some time in extending the method of
Wigner distributions to describe states of quantum systems, originally
developed for the case of continuous Cartesian coordinates and momenta 
\cite{Wey}-\cite{GM}, to the case of finite-dimensional quantum systems 
\cite{Woo}-\cite{AD}.

In its original version, the Wigner distribution is a function on the
classical phase space, real but not necessarily  pointwise nonnegative,
which describes completely any pure or mixed quantum state. Even though it
cannot be interpreted as a probability distribution on phase space, it does
lead to the correct marginal position and momentum probability distributions
as determined by quantum mechanics.

Among the early efforts to set up Wigner distribution for states of quantum
systems with a finite-dimensional state space, one may mention the work of
Feynman and of Wootters \cite{Woo}. In the former, attention was devoted 
to the two-dimensional case, drawing on the treatment of spin in quantum 
mechanics.
In the latter it was shown that one has to treat separately the cases where
the dimension of the state space is a power of two, and those where it is odd.
In the odd case one has to handle first the case of odd prime dimension, and
then pass to the general odd situation by forming a Cartesian product \ of the
prime cases.

The approach of Jagannathan \cite{Woo} on the other hand is based on the
Weyl-ordered unitary operators for translations on a phase-space lattice,
leading to the discrete Wigner distribution through the associated
characteristic function. The more recent independent work of Luis and
Pe\v rina \cite{LP} uses a similar approach, but presents a thorough
analysis of the problem.

In a previous paper \cite{CE} it has been shown that one can arrive 
at the Wigner
distribution concept, in the case of continuous variables, by a novel route
starting from an idea of Dirac \cite{Dir} to describe a general 
quantum-mechanical operator by a collection of mixed matrix elements,  using vectors chosen from two different orthonormal bases in Hilbert space. 
The steps that lead from Dirac's starting point to the expression for the 
Wigner distribution, indeed even the introduction of classical phase space 
ideas to describe operators, are particularly transparent and elementary, and
they automatically ensure the property of correctly reproducing the 
quantum-mechanical probability distributions as marginals.

The purpose of the present paper is to show that the same approach based on
Dirac's method can be used in the finite-dimensional case as well to set up
the Wigner distribution formalism, incorporating all the desirable features
including the reproduction of the marginals. It is worth particularly
emphasizing that, denoting the dimensionality of the state space by $N$, the
present approach works uniformly for all $N$; there is no need to treat
separately the cases of $N$ a power of two, $N$ an odd prime and $N$ an odd
number. In the case $N=2$, the earlier results of Feynman and Wootters are
immediately recovered, without having to call upon the specific properties of
spin-$1/2$ systems.

The construction presented here assumes particular significance in the light
of the intimate connection between Wigner distributions and mutually unbiased
bases \cite{Iva}  as was brought out by Wootters and coworkers in a
series of insightful papers \cite{Gib}. Mutually unbiased bases, 
in turn, are known to be related to questions pertaining to 
 affine planes in finite geometries, mutually orthogonal arrays, 
complex polytopes and finite designs \cite{SP} and one expects 
that the work presented here would provide  a new mathematical perspective to 
some of these questions and their interrelations.   

A brief summary of the present work is as follows. In Section 2 we discuss
the kinematics of N-level quantum systems. In particular we 
examine the trace of product of two operators ${\widehat A}$ and 
${\widehat B}$ and show how this can be
expressed as a phase space sum of products of mixed matrix
elements of the operators involved such that it manifestly reflects the  
symmetry under interchange of ${\widehat A}$ and ${\widehat B}$ . 
This entails introducing a kernel whose properties are
investigated in Section 3. In Section 4, we show  how, 
by  taking the `square root' of this kernel in a certain fashion,  one is 
directly led to the concept of a Wigner distribution associated with 
operators on a $N$-dimensional Hilbert space for any $N$.  In Section 5, by
way of illustration, we consider the case of a single qubit and recover known
results with  economy. Section 6 contains concluding remarks and further
outlook.

\section{Kinematics of an $N$-level quantum system}

We consider a quantum system possessing $N$ independent states, so that its
state space is a complex (finite-dimensional) Hilbert space $\mathcal{H}%
^{\left(  N\right)  }$ of (complex) dimension $N$. We select a particular
orthonormal basis for \ $\mathcal{H}^{\left(  N\right)  }$, written as
$|q\rangle$, with $q=0,1,...,N-1$, to be called the set of ``position
eigenstates" of the system. Then:%
\begin{equation}
\eqalign{ \langle q|q^{\prime}\rangle =\delta_{q,q^{\prime}},\; q,q^{\prime}=0,1,...,N-1; \\%
{\displaystyle\sum\limits_{q=0}^{N-1}}
|q\rangle\langle q|=\mathbb{I} .}
\end{equation}
A general vector $|\psi\rangle\in\mathcal{H}^{\left(  N\right)  }$ is
described in this basis by a corresponding wave function which is an
$N$-component complex column vector:%
\begin{equation}
\eqalign{
\psi\left(  q\right)  =\langle q|\psi\rangle ,\\
\langle \psi|\psi\rangle =\left\Vert \psi\right\Vert ^{2}=%
{\displaystyle\sum\limits_{q=0}^{N-1}}
\langle \psi|q\rangle \langle q|\psi\rangle =%
{\displaystyle\sum\limits_{q=0}^{N-1}}
\left\vert \psi\left(  q\right)  \right\vert ^{2}.}
\end{equation}

By means of an $N$-point Fourier series transformation we arrive at a
complementary orthonormal basis of ``momentum eigenstates" $|p\rangle$ with
$p=0,1,..., \mbox{$N-1$}$. The principal equations are:%
\begin{equation}
\eqalign{ 
|p\rangle = \frac{1}{\sqrt{N}} {\displaystyle\sum_{q=0}^{N-1}} 
e^{2\pi iqp/N} ~|q\rangle,\\
\langle p|p^{\prime}\rangle =\delta_{p,p^{\prime}},\; p,p^{\prime}=0,1,...,N-1, \\
{\displaystyle\sum\limits_{p=0}^{N-1}}
|p\rangle\langle p|=\mathbb{I},\\
\langle q|p\rangle =\frac{1}{\sqrt{N}}e^{2\pi iqp/N}. }
\end{equation}

Now consider a general operator $\widehat{A}$ on $\mathcal{H}^{\left(
N\right)  }$.  Using either the basis $\left\{  |q\rangle\right\}  $ or the
basis $\left\{  |p\rangle\right\}  $ for $\mathcal{H}^{\left(  N\right)  }$,
it can be completely described by the corresponding $N\times N$ square complex
matrices  $\langle q^{\prime}|\widehat{A}|q \rangle $ or
 $\langle p^{\prime}|\widehat{A}|p\rangle $. Following the method
of Dirac, however, we can equally well describe $\widehat{A}$ completely by
the collection of  ``mixed matrix elements" $\langle q|\widehat
{A}|p\rangle $; we call this an $N\times N$ ``array" rather than a matrix
 since operator multiplication is not simply the multiplication of these
arrays thought of as matrices. We also notice that with the introduction of such arrays the  step to a
``phase-space" description of  $\widehat{A}$ has been taken. More precisely, we define the (left) phase-space representative
of  $\widehat{A}$ by:%
\begin{equation}
\fl A_{l}\left(  q,p\right)  =\langle q|\widehat{A}|p\rangle
\langle p|q\rangle 
=Tr\left\{  \widehat{A}|p\rangle\langle p|q\rangle\langle q|\right\}
=\frac{1}{\sqrt{N}}\langle q|\widehat{A}|p\rangle \exp\left\{
-2\pi ipq/N\right\}.
\end{equation}
(By interchanging the r\^{o}les of $q$ and $p$ we can equally well define an
expression $A_{r}\left(  q,p\right)=\langle
p|\widehat{A}|q\rangle \langle q|p\rangle   $, however we will work with the
quantities $A_{l}\left(  q,p\right)  $). The following are immediate
consequences of this definition:
\begin{equation}
\eqalign{
{\displaystyle\sum\limits_{p}}
A_{l}\left(  q,p\right)  =\langle q|\widehat{A}|q\rangle ,\\%
{\displaystyle\sum\limits_{q}}
A_{l}\left(  q,p\right)  =\langle p|\widehat{A}|p\rangle ,\\%
{\displaystyle\sum\limits_{q,p}}
A_{l}\left(  q,p\right)  =Tr\left\{  \widehat{A}\right\} .}
\label{sums}%
\end{equation}

We may notice at this point that even for hermitian $\widehat{A}$,
$A_{l}\left(  q,p\right)  $ is in general complex.

Now take two operators $\widehat{A}$ and $\widehat{B}$ and the trace of their
product. We can express this in terms of their (left) phase-space
representatives as follows:%
\begin{eqnarray}
Tr\left\{  \widehat{A}\widehat{B}\right\} & = &
N {\displaystyle\sum_{q,p}} A_\ell(q,p)B_r(q,p)=
{\displaystyle\sum\limits_{q,p}}
{\displaystyle\sum\limits_{q^{\prime},p^{\prime}}}
\langle q|\widehat{A}|p\rangle \langle p|q^{\prime
}\rangle \langle q^{\prime}|\widehat{B}|p^{\prime}\rangle
\langle p^{\prime}|q\rangle \nonumber \\
&=& {\displaystyle\sum\limits_{q,p}}
{\displaystyle\sum\limits_{q^{\prime},p^{\prime}}}
A_{l}\left(  q,p\right)  K_{l}\left(  q,p;q^{\prime},p^{\prime}\right)
B_{l}\left(  q^{\prime},p^{\prime}\right),
\label{trace3}%
\end{eqnarray}
where:%
\begin{equation}%
\fl 
K_{l}\left(  q,p;q^{\prime},p^{\prime}\right)  =N^{2}\langle
q|p\rangle \langle p|q^{\prime}\rangle \langle
q^{\prime}|p^{\prime}\rangle \langle p^{\prime}|q\rangle 
=\exp\left\{  2\pi i\left(  q-q^{\prime}\right)  \left(  p-p^{\prime}\right)
/N\right\}.
\label{kernel1}%
\end{equation}
Thus an important phase-space kernel $K_{l}$ has been introduced. We note in
passing that (apart from the $N^{2}$ factor) it is a four-vertex Bargmann
invariant, so its phase is an instance of the kinematic geometric phase \cite{MS}.

The study of the detailed properties of $K_{l}$ will lead us to the solution
of setting up a physically reasonable Wigner distribution, \textit{for any
value of the dimension }$N$.

\section{Properties of the Kernel $K_{l}$}

We can regard $K_{l}\left(  q,p;q^{\prime},p^{\prime}\right)  $ as defined in
Eq. (\ref{kernel1}) as constituting a complex square matrix of dimension
$N^{2}$, with the first pair of arguments $\left(  q,p\right)  $ being row
index and the second pair $\left(  q^{\prime},p^{\prime}\right)  $  column
index\footnote{We introduce below a more compact efficient notation to express
this.}. We denote by $\mathcal{K}^{\left(  N^{2}\right)}$ a complex linear
space of dimension $N^2$, made up of vectors $f$ with components $f(q,p)$:
\begin{equation}
f~\in~ \mathcal{K}^{\left(  N^{2}\right)} \rightarrow f(q,p), q,p=0,1,2\cdots,
N-1.
\label{31}
\end{equation}
It is to be understood that these vectors are ``periodic''
in the sense that 
\begin{equation}
f(q+nN,p+n^\prime N)=f(q,p), n,~n^\prime =0,~\pm 1,~\pm 2,\cdots  .
\label{32}
\end{equation}
The norm is defined in the natural way by 
\begin{equation}
||f||^2=(f,f)=\displaystyle\sum_{q,p=0,1,\cdots}^{N-1}~|f(q,p)|^2.
\label{33}
\end{equation}
Then $K_{l}$ acts on such vectors according to 
\begin{equation}
\left(K_lf\right)\left(q,p\right)=\displaystyle\sum_{q^\prime,p^\prime}
K_{l}\left(  q,p;q^{\prime},p^{\prime}\right)~f(q^\prime,p^\prime).
\label{34}
\end{equation}
The following properties are immediately evident:

\begin{itemize}
\item Symmetry:%
\begin{equation}
K_{l}\left(  q,p;q^{\prime},p^{\prime}\right)  =K_{l}\left(  q^{\prime
},p^{\prime};q,p\right) ; \label{kernel2}%
\end{equation}

\item Essential unitarity:%
\begin{equation}%
{\displaystyle\sum\limits_{q^{\prime},p^{\prime}}}
K_{l}\left(  q,p;q^{\prime},p^{\prime}\right)  K_{l}\left(  q^{\prime\prime
},p^{\prime\prime};q^{\prime},p^{\prime}\right)  ^{\ast}=N^{2}\delta
_{qq^{\prime\prime}}\delta_{pp^{\prime\prime}} ;\label{kernel3}%
\end{equation}

\item Translation invariance:%
\begin{eqnarray}
K_{l}\left(  q+q_0,p+p_0;q^{\prime}+q_0,p^{\prime}+ p_0\right)  
=K_{l}\left(q,p; q^{\prime},p^{\prime}\right), &&\label{kernel2a}\\
~~~~~~~~~~~~~~~~~~~~~~~~~~~~~~~~~q_0,p_0=0,1,2,\cdots,N-1.  &&\nonumber
\end{eqnarray}

\end{itemize}
Here and in the following we interpret translated arguments
$q+q_0,p+p_0,\cdots$ as always taken modulo $N$, so that they always lie in the
range $0,1,\cdots,N-1$. Property (\ref{kernel3}) means that any eigenvalue of 
$K_l$ is of the form $Ne^{i\varphi}$ for some phase $\varphi$. In addition to the
above, the following `marginals' properties are also evident from the
definition (\ref{kernel1}):
\numparts
\begin{eqnarray}
&& {\displaystyle\sum\limits_{p^{\prime}}}
K_{l}\left(  q,p;q^{\prime},p^{\prime}\right)  =N\delta_{qq^{\prime}},\; {\rm independent} \; of\; p , \label{kernel4a} \\
&& {\displaystyle\sum\limits_{q^{\prime}}}
K _{l}\left(  q,p;q^{\prime},p^{\prime}\right)  =N\delta_{pp^{\prime}},\; {\rm independent} \;of\;q . \label{kernel4b}
\end{eqnarray}
\endnumparts
These are particularly important for the Wigner
distribution problem, so we explore them in some detail and relate them to
the eigenvalue and eigenvector properties of $K_{l}$. From either one of Eqs.(\ref{kernel4a},\ref{kernel4b}) we get the
(weaker) relations:%
\begin{equation}%
{\displaystyle\sum\limits_{q^{\prime}p^{\prime}}}
K_{l}\left(  q,p;q^{\prime},p^{\prime}\right)  =N, \; {\rm independent \; of}\;q,p. \label{kernel5}%
\end{equation}

Let us introduce a single symbol $\sigma$ to denote the pair $\left(
q,p\right)  $ by the definition:%
\begin{equation}
\sigma=qN+p+1.
\end{equation}
Thus $\sigma$ runs from $1$ to $N^{2}$: for $q=0,p=0,1,...,N-1$ we have
$\sigma=1,2,...,N$; for $q=1,p=0,1,...,N-1$ we have $\sigma=N+1,N+2,...,2N$;
and so on. For summations and Kronecker symbols we have the rules:
\begin{eqnarray}
{\displaystyle\sum\limits_{\stackrel{p}{q\;fixed}}} \cdots &=&
{\displaystyle\sum\limits_{\sigma=qN+1,qN+2,...,\left(  q+1\right)  N }}
\cdots,\\
{\displaystyle\sum\limits_{\stackrel{q}{p\;fixed}}}
\cdots &=&
{\displaystyle\sum\limits_{sigma=p+1,N+p+1,2N+p+1,...,\left(  N-1\right)
N+p+1 }} \cdots \nonumber,\\
{\displaystyle\sum\limits_{qp}} \cdots &=& ~~
{\displaystyle\sum\limits_{\sigma=1}^{N^{2}}} \cdots ,\nonumber \\
~~~\delta_{\sigma\sigma^{\prime}}&=&~~\delta_{qq^{\prime}}\delta_{pp^{\prime}}. \nonumber
\end{eqnarray}
We hereafter use $\sigma$ or $q,p$ interchangeably as convenient. 
The kernel $K_{l}\left(  q,p;q^{\prime},p^{\prime}\right)  $ can now be
written as $K_{l}\left(  \sigma;\sigma^{\prime}\right)  $, while vectors  
$f\in \mathcal{K}^{\left(N^{2}\right)  }$ have components $f(\sigma)$. 
In addition to the properties (\ref{kernel2}),(\ref{kernel3}),(\ref{kernel4a},\ref{kernel4b}),(\ref{kernel5}) we have the trace property following from
(\ref{kernel1}):%
\begin{equation}
TrK_{l}=%
{\displaystyle\sum\limits_{\sigma}}
K_{l}\left(  \sigma,\sigma\right)  =N^{2}.%
\end{equation}

With this notation one can now see that the marginals properties
(\ref{kernel4a},\ref{kernel4b}) can be expressed as follows. For each $q^{\prime
}=0,1,...,N-1$ we define a vector $U_{q^{\prime}}$ in $\mathcal{K}%
^{\left(  N^{2}\right)  }$, forming altogether a set of $N$ real orthonormal
vectors (not a basis!) by:%
\begin{equation}%
\eqalign{
U_{q^{\prime}}\left(  \sigma\right)  =\frac{1}{\sqrt{N}}%
\delta_{qq^{\prime}}, \;{\rm independent~ of}\; p,\\
(U_{q^{\prime}},U_{q})=\delta_{q^{\prime}q}. }
\label{uqu}%
\end{equation}
Then Eq. (\ref{kernel4a}) translates exactly into the statement:%
\begin{equation}
K_{l}U_{q}=NU_{q},\;q=0,1,...,N-1.
\end{equation}
Similarly, for each $p^{\prime}=0,1,...,N-1$ we define a vector
$V_{p^{\prime}}$ in  $\mathcal{K}^{\left(  N^{2}\right)  }$ forming
altogether a set of $N$ real orthonormal vectors (again, not a basis!) by:
\begin{equation}%
\eqalign{
V_{p^{\prime}}\left(  \sigma\right)  =\frac{1}{\sqrt{N}}%
\delta_{pp^{\prime}},\;{\rm independent~ of}\; q,\\
(V_{p^{\prime}},V_{p})=\delta_{p^{\prime}p}. }
\label{vupi}%
\end{equation}
Then Eq. (\ref{kernel4b}) translates into:%
\begin{equation}
K_{l}V_{p}=NV_{p},\;p=0,1,...,N-1.
\end{equation}

These real eigenvectors $U_{q}$ and $V_{p}$ are mutually nonorthogonal:%
\begin{equation}
(V_{p},U_{q})=
{\displaystyle\sum\limits_{\sigma^{\prime}}}
V_{p}\left(  \sigma^{\prime}\right)  U_{q}\left(  \sigma^{\prime}\right)
=\frac{1}{N}.%
\end{equation}
This leads to the single linear dependence relation among the $2N$ (real)
vectors $U_{q},V_{p}$:%
\begin{equation}
{\displaystyle\sum\limits_{q}}\,
U_{q}= {\displaystyle\sum\limits_{p}}\,
V_{p},%
\end{equation}
which can also be read off from Eqs. (\ref{uqu}) and (\ref{vupi}). Therefore,
the $U_{q}$'s and $V_{p}$'s together span an $\left(  2N-1\right)  $-dimensional
subspace $\mathcal{K}^{\left(  2N-1\right)  }$ in $\mathcal{K}^{\left(
N^{2}\right)  }$, over which $K_{l}$ reduces to $N$ times the identity.
We can construct an orthonormal basis of $\left(
2N-1\right)  $ real vectors for $\mathcal{K}^{\left(  2N-1\right)  }$ for
instance by the following recipe:%
\begin{equation}
\eqalign{
\Psi_{0}=\frac{1}{\sqrt{N}}
{\displaystyle\sum\limits_{q}}\,
U_{q}=\frac{1}{\sqrt{N}}
{\displaystyle\sum\limits_{p}} \, V_{p},\\
\widetilde{U}_{j}=\frac{1}{\sqrt{j\left(  j+1\right)  }}\left(  U_{0}%
+U_{1}+...+U_{j-1}-jU_{j}\right)  ,\;j=1,2,...,N-1,\\
\widetilde{V}_{j}=\frac{1}{\sqrt{j\left(  j+1\right)  }}\left(  V_{0}%
+V_{1}+...+V_{j-1}-jV_{j}\right)  ,\;j=1,2,...,N-1,\\
(\widetilde{U}_{j^{\prime}},\widetilde{U}_{j})=(\widetilde{V}_{j^{\prime}},
\widetilde{V}_{j})=\delta_{j^{\prime}j},\;(\Psi_{0},\Psi_{0})=1,\\
(\widetilde{U}_{j^{\prime}},\Psi_{0})=(\widetilde{V}_{j^{\prime}},\Psi
_{0})=(\widetilde{U}_{j^{\prime}},\widetilde{V}_{j})=0. }
\label{basis1}%
\end{equation}

If the orthogonal complement of $\mathcal{K}^{\left(  2N-1\right)  }$ in
$\mathcal{K}^{\left(  N^{2}\right)  }$, of dimension $\left(  N-1\right)
^{2}$, is written as $\mathcal{K}^{\left(  N-1\right)  ^{2}}$, i.e.:%
\begin{equation}
\mathcal{K}^{\left(  N^{2}\right)  }=\mathcal{K}^{\left(  2N-1\right)  }%
\oplus\mathcal{K}^{\left(  N-1\right)  ^{2}}, \label{decomposition}%
\end{equation}
then we can supplement the basis (\ref{basis1}) for $\mathcal{K}^{\left(
2N-1\right)  }$ by (any) additional real orthonormal vectors to span
$\mathcal{K}^{\left(  N-1\right)  ^{2}}$. The essential unitarity of $K_{l}$
 means that it leaves $\mathcal{K}^{\left(  N-1\right)  ^{2}}$ also
invariant; the transition from the original (standard) basis of $\mathcal{K}%
^{\left(  N^{2}\right)  }$ to the present one can be accomplished  by an
element of the real orthogonal rotation group $SO\left(  N^{2}\right)  $, thus
preserving the symmetry (\ref{kernel2}) of $K_{l}$.  Therefore the matrix
$K_{l}$ has the following structure in a (real) basis adapted to the
decomposition (\ref{decomposition}):%
\begin{equation}
K_{l}\rightarrow\left(
\begin{array}
[c]{cc}%
\mathbf{N\cdot}\mathbb{I} & \mathbf{0}\\
\mathbf{0} & \mathbf{A}+i\mathbf{B}%
\end{array}
\right). \label{kappael}%
\end{equation}
The unit matrix is of dimension $\left(  2N-1\right)  $, while the two real
$\left(  N-1\right)  ^{2}$-dimensional  matrices $\mathbf{A}$ and
$\mathbf{B}$ obey:%
\begin{equation}
\eqalign{
\mathbf{A}^{T}=\mathbf{A,}\;\mathbf{B}^{T}=\mathbf{B,}\; \mathbf{AB=BA},\\
\mathbf{A}^{2}+\mathbf{B}^{2}=N^{2}\cdot\mathbb{I}_{\left(  N-1\right)
^{2}\times\left(  N-1\right)  ^{2}},\\
Tr\{\mathbf{A}\}=-N\left(  N-1\right)  ,\;Tr\{\mathbf{B}\}=0 .}
\label{prop}%
\end{equation}
Thus the matrix  $\mathbf{A}+i\mathbf{B}$  can definitely be diagonalized
 by a real rotation in $\left(  N-1\right)  ^{2}$ dimensions, i.e., by an
element of $SO\left(  \left(  N-1\right)  ^{2}\right)  $, and each eigenvalue
of  $\mathbf{A}+i\mathbf{B}$  is of the form $Ne^{i\varphi}$  for some
angle  $\varphi$.

It now turns out that we can carry through this diagonalisation process
explicitly. The translation invariance (\ref{kernel2a}) of $K_l$ means that
the eigenvectors of $K_l$ can be constructed as ``plane waves'' in phase
space. We can obtain a complete real orthonormal set of vectors of $K_l$ in 
$\mathcal{K}^{\left(  N^{2}\right)}$ by this route, recovering 
the subset of eigenvectors (\ref{basis1}) as part of a complete set. 

For each point $\sigma_{0} =(q_0,p_0)$ we define a unit vector $\chi_{\sigma_{0}}$ with components 
\begin{equation}
\chi_{\sigma_{0}}(\sigma)=\frac{1}{N} \exp(2\pi i(q_0p+p_0q)/N)
\label{323}
\end{equation}
(we see that condition (\ref{32}) is indeed obeyed). Thus we have exactly
$N^2$ vectors $\chi_{\sigma_{0}}$. Using the modulo $N$ rule for phase space
arguments we then easily obtain the following: 
\numparts
\begin{eqnarray} 
K_l\chi_{\sigma_{0}} &=& N  e^{-2\pi iq_0 p_0/N}\chi_{\sigma_0}, \\
\left(\chi_{\sigma_{0}^\prime}, \chi_{\sigma_{0}}\right) \label{324a}&=& 
\delta_{\sigma_{0}^\prime~\sigma_{0}}.
\label{324b}
\end{eqnarray}
\endnumparts
Therefore we have achieved full diagonalisation of  $K_l$, with  
$\{\chi_{\sigma_{0}}\}$ forming an orthonormal basis in  
$\mathcal{K}^{\left(  N^{2}\right)}$. The previously found (real) basis for
the subspace $\mathcal{K}^{\left(2N-1\right)}$, made up exclusively of
eigenvectors of $K_l$ with eigenvalues $N$, is essentially the subset of
$(2N-1)$ vectors $ \chi_{q_0,0}$ for $q_0=0,1,\cdots,N-1$ and
$\chi_{0,p_0}$ for $p_0= 1,\cdots,N-1$. Indeed we find 
\begin{equation}
\eqalign{ U_q = \frac{1}{\sqrt{N}} \displaystyle\sum_{p_0=0}^{N-1} e^{-2\pi iqp_0/N}
\chi_{0,p_0},\nonumber\\
V_p =  \frac{1}{\sqrt{N}} \displaystyle\sum_{q_0=0}^{N-1} e^{-2\pi iq_0p/N}
\chi_{q_0,0}. }
\label{325}
\end{equation}
The remaining $(N-1)^2$ eigenvectors $\chi_{\sigma_0}$ for 
$q_0,p_0=1,2,\cdots,N-1$ span the orthogonal subspace 
$\mathcal{K}^{\left(  {N-1}^{2}\right)}$. Here we have in detail the following
structure. The two eigenvectors $\chi_{q_0,p_0}$ and $\chi_{N-q_0,N-p_0}$ are 
degenerate, and their components are related by complex conjugation : 
\begin{eqnarray}
K_l \chi_{q_0,p_0} &=& N e^{-2\pi iq_0p_0/N} \chi_{q_0,p_0},\nonumber\\
K_l \chi_{N-q_0,N-p_0} &=& N e^{-2\pi iq_0p_0/N} \chi_{N-q_0,N-p_0};\label{326}\\
\chi_{N-q_0,N-p_0}(\sigma) &=&\chi_{q_0,p_0}(\sigma)^*.
\nonumber
\end{eqnarray} 
Therefore we have a pattern that depends on the parity of $N$. For odd $N$, we
have $(N-1)^2/2$ distinct degenerate pairs of mutually complex conjugate
orthogonal eigenvectors $ \{\chi_{q_0,p_0}, \chi_{N-q_0,N-p_0}\}$ for $
q_0=1,2,\cdots,(N-1)/2$ and $p_0=1,2,\cdots,N-1$. For even $N$ we have one
real eigenvector $\chi_{N/2,N/2}$ with eigenvalue $N(-1)^{N/2}$ followed by $ 
((N-1)^2-1)/2$ distinct degenerate pairs 
$\{\chi_{q_0,p_0},\chi_{N-q_0,N-p_0}\}$ where we omit $q_0=p_0=N/2$. In either
case it is clear that by passing to the real and imaginary parts of 
$\chi_{q_0,p_0}$, while leaving $\chi_{N/2,N/2}$ unchanged, we get a real
orthonormal basis for $\mathcal{K}^{\left(N-1\right)^2}$ in which the matrix 
$\mathbf{A}+i\mathbf{B}$ of Eq. (\ref{kappael}) is diagonal. 

Equipped with these important properties of $K_{l}$ we turn to
Eq. (\ref{trace3})  from where we can find the route to the Wigner distribution.

\section{The Kernel $\xi$ and the Wigner Distribution}

Motivated by the structure (\ref{trace3}) for $Tr\left\{  \widehat{A}%
\widehat{B}\right\}  $ for two general operators $\widehat{A}$ and
$\widehat{B}$ on $\mathcal{H}^{\left(  N\right)  }$, we try to express the
kernel \ $K_{l}\left(  q,p;q^{\prime},p^{\prime}\right)  $ in the form:%
\begin{equation}
K_{l}\left(  \sigma,\sigma^{\prime}\right)  =
{\displaystyle\sum\limits_{\sigma^{\prime\prime}}}
\xi\left(  \sigma^{\prime\prime},\sigma\right)  \xi\left(  \sigma
^{\prime\prime},\sigma^{\prime}\right), \label{41}
\end{equation}
with suitable conditions imposed on $\xi$.  The desirable conditions are, as
with $K_{l}$ itself: symmetry, essential unitarity, translation invariance and marginal conditions similar to Eqs. (\ref{kernel4a},\ref{kernel4b}) for $K_{l}$:
\numparts
\begin{eqnarray}
&& \xi\left(  \sigma,\sigma^{\prime}\right)
=\xi\left(  \sigma^{\prime},\sigma\right),  \label{eigen2a}\\
&& {\displaystyle\sum\limits_{\sigma^{\prime}}} \, 
\xi\left(  \sigma,\sigma^{\prime}\right)  \xi\left(  \sigma^{\prime\prime
},\sigma^{\prime}\right)  ^{\ast}=N\delta_{\sigma\sigma^{\prime\prime}} \label{eigen2b},\\
&& \xi(q+q_0,p+p_0;q^\prime +q_0,p^\prime +p_0)= \xi(q,p;q^\prime ,p^\prime)
\label{eigen2c},\\
&& \xi \, U_{q}=\sqrt{N}U_{q},\;\xi \, V_{p}
=\sqrt{N}V_{p}.
\label{eigen2d}%
\end{eqnarray}
\endnumparts
Here we have expressed the last marginals conditions already in terms of the
eigenvectors (not all independent!) $U_{q},V_{p}$ of $K_{l}$ lying in 
$\mathcal{K}^{\left(2N-1\right)}$. More explicitly they read 
\begin{equation}
\eqalign{
\displaystyle\sum_{p^\prime} \xi(q,p;q^\prime ,p^\prime)= 
\sqrt{N}\delta_{qq^\prime},\\
\displaystyle\sum_{q^\prime} \xi(q,p;q^\prime ,p^\prime)=
\sqrt{N}\delta_{pp^\prime}. }
\label{43}
\end{equation}
The detailed analysis of the eigenvectors and eigenvalues of $K_l$ in the
previous section immediately leads to solutions for $\xi$. The translation
invariance of (\ref{eigen2c}) is ensured by arranging that the ``plane waves''
eigenvectors $\chi_{q_0,p_0}$ of $K_l$ are eigenvectors of $\xi$ as well. We
take $\xi$ to obey:
\numparts
\begin{eqnarray}
\xi~\chi_{q_0,0}=\sqrt{N}\chi_{q_0,0},&& q _0=0,1,\cdots,N-1;\label{44a}\\\xi~\chi_{0,p_0}=\sqrt{N}\chi_{0,p_0},&& p_0=0,1,\cdots,N-1;
\label{44b}\\
\xi~\left(\chi_{q_0,p_0}~ {\rm or}~\chi_{N-q_0,N-p_0}\right) &=& \pm
\sqrt{N}~e^{-i\pi q_0p_0/N}\left(\chi_{q_0,p_0}~ {\rm or}~\chi_{N-q_0,N-p_0}\right), \label{44c}\\ 
&~& q_0,p_0=1,\cdots,N-1. \nonumber
\end{eqnarray} 
\endnumparts
In the subspaces $\mathcal{K}^{\left(2N-1\right)}$ and 
$\mathcal{K}^{\left({N-1}^2\right)}$ we then have:
\begin{equation}
\eqalign{
\xi = \sqrt{N}\,\cdot  \mathbb{I} &~~ {\rm on} ~
\mathcal{K}^{\left(2N-1\right)},
\\ 
\xi = \left(\mathbf{A}+i\mathbf{B}\right)^{1/2} &~~ {\rm on}~
\mathcal{K}^{\left(N-1\right)^2}. }
\label{45}
\end{equation} 
Equations (\ref{44a}) ensure the validity of the marginals properties 
(\ref{eigen2d}) or (\ref{43}) while Eq. (\ref{eigen2b}) is obeyed by 
construction. It is the symmetry requirement (\ref{eigen2a}) that dictates
that in the case of degenerate orthonormal pairs of $K_l$ eigenvectors  
$ \{\chi_{q_0,p_0}, \chi_{N-q_0,N-p_0}\}$ we choose the square root of the
eigenvalue $N e^{-2\pi iq_0p_0/N}$ of $K_l$ in the same way; this is expressed 
in Eq.  (\ref{44c}). Thus we see: for odd $N$ there is a $2^{(N-1)^2/2}$- 
fold freedom in the choice of $\xi$; for $N$ even there is a 
$2^{((N-1)^2+1)/2}$-fold freedom. In each case, a particular square root of 
$\mathbf{A}+i\mathbf{B}$ is involved in (\ref{45}). 
 
With any such $\xi$, we can return to Eq. (\ref{trace3}) and write it in a
manifestly kernel-independent manner:
\begin{equation}
Tr\left\{  \widehat{A}\widehat{B}\right\}  =N
{\displaystyle\sum\limits_{q,p}}
A\left(  q,p\right)  B\left(  q,p\right),  \label{trace4}%
\end{equation}
where:%
\begin{eqnarray}
A\left(  q,p\right) & =&\frac{1}{\sqrt{N}}
{\displaystyle\sum\limits_{q^{\prime},p^{\prime}}}
\xi\left(  q,p;q^{\prime},p^{\prime}\right)  A_{l}\left(  q^{\prime}%
,p^{\prime}\right)  \label{kernel6}\\
&=&\frac{1}{\sqrt{N}}
{\displaystyle\sum\limits_{q^{\prime},p^{\prime}}}
\xi\left(  q,p;q^{\prime},p^{\prime}\right)  \langle q^{\prime}
|\widehat{A}|p^{\prime}\rangle \langle p^{\prime}|q^{\prime
}\rangle ,\nonumber
\end{eqnarray}
with a similar expression for $B\left(  q,p\right)  $ in terms of
 $\widehat{B}$. We will show below that for hermitian $\widehat{A}$, 
$A(q,p)$ is real. Combining Eqs.(\ref{sums}) and (\ref{eigen2d}) we have ensured the marginals properties:%
\begin{equation}
\eqalign{
{\displaystyle\sum\limits_{p}}
A\left(  q,p\right)  =\langle q|\widehat{A}|q\rangle, \\
{\displaystyle\sum\limits_{q}}
A\left(  q,p\right)  =\langle p|\widehat{A}|p\rangle.}
\label{sums2}%
\end{equation}

For the density matrix $\widehat{\rho}$ describing some pure or mixed state of the $N$-level system, we then have the real Wigner distribution:%
\begin{equation}
W\left(  q,p\right)  =\frac{1}{\sqrt{N}}
{\displaystyle\sum\limits_{q^{\prime},p^{\prime}}}
\xi\left(  q,p;q^{\prime},p^{\prime}\right)  \langle q^{\prime}%
|\widehat{\rho}|p^{\prime}\rangle \langle p^{\prime}|q^{\prime
}\rangle \label{realWigner}%
\end{equation}
and by Eqs.(\ref{sums2}) the two marginal probability distributions in $q$ and
$p$ are immediately recovered. In particular, we find that for position 
eigenstates and momentum eigenstates the freedom in the choice of $\xi$ 
( which in any case is limited to its action on 
$\mathcal{K}^{\left(  {N-1}^{2}\right)}$) does not matter and we get the  anticipated results:%
\begin{equation}
\eqalign{
\widehat{\rho}=|q^{\prime}\rangle\langle q^{\prime}|\;\Rightarrow
\;W\left(  q,p\right)  =\frac{1}{N}\delta_{qq^{\prime}},\; {\rm independent~ of~ }p,\\
\widehat{\rho}=|p^{\prime}\rangle\langle p^{\prime}|\;\Rightarrow
\;W\left(  q,p\right)  =\frac{1}{N}\delta_{pp^{\prime}}, \; {\rm independent~ of~}q. }
\label{realWigner2}%
\end{equation}

Returning to the Wigner distribution (\ref{realWigner}) we may rewrite it as 
\begin{equation}
W\left(q,p\right)= \frac{1}{N} Tr\{ \rho {\widehat W}(q,p)\}
\label{411}
\end{equation}
by introducing elements of Wigner basis \cite{Pra} or phase point
operators \cite{Fir}: 
\begin{equation}
{\widehat W}\left(  q,p\right)  =\sqrt{N} 
\sum_{q^\prime,p^\prime} \xi\left(  q,p;q^{\prime},p^{\prime}\right)   
\langle p^{\prime}|q^{\prime}\rangle 
|p^{\prime}\rangle \langle q^{\prime}|.
\label{dd}
\end{equation}
It is an interesting exercise to verify, by combining the definition 
(\ref{kernel1}) of $K_l$ and (\ref{41},\ref{eigen2b}), that these are
hermitian:
\begin{equation}
\widehat{W}(q,p)^\dagger =\widehat{W}(q,p).
\end{equation}
This proves that $W(q,p)$, and more generally 
$A(q,p)=Tr\{\widehat{A} \widehat{W}(q,p)\}$ for hermitian $\widehat{A}$, are
both real. In addition one can check, by virtue of eqs (\ref{41},\ref{eigen2a}$-d$), they satisfy  
\begin{equation}
\eqalign{
Tr \{{\widehat W}\left(\sigma \right)\} = 1 ,\\
Tr  \{{\widehat W}\left(\sigma \right){\widehat W}
\left(\sigma^\prime \right)\}  = N \delta_{\sigma \sigma^{\prime}}.}
\label{414}
\end{equation}

\section{The case of $N=2$: the Qubit}

This case is particularly interesting in that earlier treatments have had to
treat it on its own, in a sense in an ad hoc manner, as distinct from $N$ an
odd prime or an odd integer. In the standard basis for the two-dimensional Hilbert space $\mathcal{H}^{\left(  2\right)  }$ made up of $|q\rangle$ for
$q=0,1$, accompanied by its complementary basis $|p\rangle$, the matrix
$K_{l}\left(  q,p;q^{\prime},p^{\prime}\right)  $ is the following:%
\begin{equation}
K_{l}=\left(
\begin{array}
[c]{cccc}%
1 & 1 & 1 & -1\\
1 & 1 & -1 & 1\\
1 & -1 & 1 & 1\\
-1 & 1 & 1 & 1
\end{array}
\right).
\end{equation}
The rows and columns are labelled in the sequence: $\left(  q,p\right)
=\left(  0,0\right)  ,\left(  0,1\right)  ,\left(  1,0\right)  ,\left(
1,1\right)  $, and the matrix elements are read off from Eq. (\ref{kernel1}).
The three orthonormal eigenvectors of $K_{l}$ with eigenvalue $2$, spanning
the subspace $\mathcal{K}^{\left(  3\right)  }$ of the general treatment in
Section 3 are:%
\begin{equation}
\Psi_{0}=\frac{1}{2}\left(
\begin{array}
[c]{c}%
1\\
1\\
1\\
1
\end{array}
\right) ,\;\widetilde{U}_{1}=\frac{1}{2}\left(
\begin{array}
[c]{c}%
1\\
1\\
-1\\
-1
\end{array}
\right) ,\;\widetilde{V}_{1}=\frac{1}{2}\left(
\begin{array}
[c]{c}%
1\\
-1\\
1\\
-1
\end{array}
\right).
\end{equation}

We choose the fourth eigenvector of $K_{l}$, with eigenvalue  necessarily
$-2$  since $TrK_{l}=4$, to be:%
\begin{equation}
W=\frac{1}{2}\left(
\begin{array}
[c]{c}%
1\\
-1\\
-1\\
1
\end{array}
\right).
\end{equation}
Then the kernel $\xi$ can be immediately synthesized from:%
\begin{equation}
\xi\Psi_{0}=\sqrt{2}\Psi_{0},\;\xi\widetilde{U}_{1}=\sqrt{2}%
\widetilde{U}_{1},\;\xi\widetilde{V}_{1}=\sqrt{2}\widetilde{V}%
_{1},\;\xi W=i\sqrt{2}W \label{evs}
\end{equation}
and in the standard basis turns out to be:%
\begin{equation}
\xi=\frac{1}{2\sqrt{2}}\left(
\begin{array}
[c]{cccc}%
3+i & 1-i & 1-i & -1+i\\
1-i & 3+i & -1+i & 1-i\\
1-i & -1+i & 3+i & 1-i\\
-1+i & 1-i & 1-i & 3+i
\end{array}
\right). \label{52}
\end{equation}
Using the above matrix elements of $\xi$ in (\ref{dd}) we obtain,  
for the phase-point operators:%
\begin{equation} 
\begin{array}
[c]{ll}%
\widehat{\mathcal{W}}\left(  0,0\right)  =\left(
\begin{array}
[c]{cc}%
1 & \frac{1-i}{2}\\
 \frac{1+i}{2}  & 0
\end{array}
\right), & \widehat{\mathcal{W}}\left(  0,1\right)  =\left(
\begin{array}
[c]{cc}%
0 & \frac{1+i}{2}  \\
 \frac{1-i}{2} & 1
\end{array}
\right), \\
~ & ~ \\
\widehat{\mathcal{W}}\left(  1,0\right)  =\left(
\begin{array}
[c]{cc}%
1 & \frac{-1+i}{2}\\
\frac{-1-i}{2}& 0
\end{array}
\right),  & \widehat{\mathcal{W}}\left(  1,1\right)  =\left(
\begin{array}
[c]{cc}%
0 & \frac{-1-i}{2}\\
\frac{-1+i}{2} & 1
\end{array}
\right). \label{53}
\end{array}
\bigskip\end{equation}
and thereby recover the  results of Feynman and Wootters \cite{Woo} and
hence also the connection between sums of phase point operators along 
striations of the qubit phase space \cite{Gib}  and the mutually 
unbiased bases for $N=2$.

For the density operator ${\widehat \rho}= 
\frac{1}{2}(\mathbb{I}_2 +{\mathbf a}\cdot {\mathbf \sigma}),
~~{\mathbf a}\cdot {\mathbf a} \leq 1$ describing a general state of a
qubit one can easily calculate the corresponding Wigner distribution using
(\ref{realWigner}) or (\ref{411}). The results arranged in the form of a matrix 
read:
\begin{equation}
\left(\begin{array}
[c]{cc}%
1+a_1+a_2+a_3& 1+a_1-a_2-a_3\\
 1-a_1-a_2+a_3& 1-a_1+a_2-a_3
\end{array}\right).
\label{54}
\end{equation}
Using this result it is instructive to verify the validity of (\ref{trace4}) for
$ {\widehat A}= {\widehat \rho_1}= \frac{1}{2}(\mathbb{I}_2 +{\mathbf a}
\cdot {\mathbf \sigma}),$ and $ {\widehat B}= {\widehat \rho_2}= 
\frac{1}{2}(\mathbb{I}_2 
+{\mathbf b}\cdot{\mathbf \sigma})$. Further, it is easily seen from 
(\ref {54}) that  the maximum positive and maximum negative values of the 
Wigner distribution for a qubit occur when $|a_1|=|a_2|=|a_3|=1/\sqrt{3}$.

As a final remark, we notice that in calculating the kernel $\xi$, square root of $K_l$, we might have chosen $W$ in Eq.  (\ref{evs}) to be the eigenvector corresponding to the eigenvalue $-i\sqrt{2}$ instead of $+i\sqrt{2}$. This results in changing $+i$($-i$) with $-i$($+i$) in Eqs. (\ref{52},\ref{53}) and thus in an interchange of the r\^ole of $\widehat{\mathcal{W}}\left(  0,0\right)$ and $ \widehat{\mathcal{W}}\left(  0,1\right)$ with $\widehat{\mathcal{W}}\left(  1,0\right)$ and $\widehat{\mathcal{W}}\left(  1,1\right)$ respectively. Correspondingly, the coefficient $a_2$ in Eq. (\ref{54}) would change sign everywhere. This however can be of no physical consequence as is reflected
in the fact that the marginals obtained by summing over $q$ or $p$ are
independent of $a_2$.

\section{Concluding remarks.}

To conclude, we have developed a method of constructing Wigner distribution
for \mbox{$N$-level} systems which is remarkable in its directness and economy and 
works uniformly for all $N$. The construction is
entirely algebraic and solely involves finding a square root of a certain $N^2\times N^2$
complex symmetric matrix. 
No other auxiliary inputs  are required. As an illustration, we have  worked 
out the qubit case in some detail and obtained  results already known in
the literature in an extremely economic fashion. The
construction presented here provides a fresh perspective to several questions 
pertaining to quantum tomography in finite state systems and to those
associated with finite geometries which are currently being investigated with
great vigour owing to their relevance to quantum information theory. We hope
to return to some of these elsewhere.


\section*{References}

\begin{thebibliography}{99}

\bibitem{Wey} Weyl H 1927 {\it Z. Phys.} \textbf{46} 1  and 1931  {\it The Theory of Groups and Quantum Mechanics} (Dover, N.Y.) p~274

\bibitem{Wig} Wigner E P 1932 {\it Phys. Rev.} \textbf{40} 749.  For
reviews see: Hillery M, O'Connell R F, Scully M O and Wigner E P 1984 {\it Phys. Repts.} \textbf{106} 121; Kim Y S and Noz M E 1991 {\it Phase-Space Picture of Quantum Mechanics} (World Scientific, Singapore); Schleich W P 2001 {\it Quantum Optics in Phase Space} (Wiley-VCH, Weinheim)

\bibitem{GM} Groenewold H J 1946 {\it Physica} \textbf{12} 205;
Moyal J E 1949 {\it Proc. Camb. Phil. Soc.} \textbf{45} 99 

\bibitem{Woo} Jagannathan R 1976 {\it Studies in Generalized Clifford Algebras, Generalized Clifford Groups and their Physical Applications} Ph D thesis (University of Madras); Wootters W K 1987 {\it Ann. Phys. (N.Y.)} \textbf{176} 1; Feynman R P 1987 Negative Probabilities in {\it Quantum Implications. Essays
in Honour of David Bohm} Hiley B and Peat D  Eds. (Routledge, London)
 
\bibitem{Sch}  Schwinger J 1960 {\it Proc. Nat. Acad. Sci. USA}  {\bf 46} 570; Buot F A 1974 {\it    Phys. Rev. B} {\bf 10} 3700; Hannay J H and  Berry M V 1980  {\it Physica D} {\bf 1} 267; Cohen L and Scully M 1986 {\it Found. Phys.} {\bf 16} 295; Cohendet O, Combe P, Siugue M and Sirugue-Collin M 1988 {\it  J. Phys. A} {\bf 21} 2875; Galetti D and de Toledo Piza A F R 1988 {\it Physica  A} {\bf 149} 267; Kasperkovitz P and  Peev M 1994 {\it  Ann. Phys. (N. Y.)} {\bf 230} 21; Bouzouina A and Bi\`evre S 1996 {\it  Comm Math. Phys.} {\bf 178} 83; Rivas A M and Ozorio de Almeida A M 1999 {\it Ann. Phys. (N. Y.)} {\bf 276} 123 

\bibitem{VP} Vacarro J A and  Pegg D T 1990 {\it Phys. Rev. A} {\bf 41} 5156; Leonhardt U 1995 {\it Phys. Rev. Lett.} {\bf 74} 4101, 1996 {\it Phys. Rev. A}  {\bf 53} 2998  and 1996 {\it  Phys. Rev. Lett.} {\bf 76} 4293 

\bibitem{BM} Bianucci P, Miquel C, Paz J P and Saraceno M 2002 {\it  Phys. Lett. A} {\bf 297} 353; Miquel C, Paz J P, Saraceno M, Knill E, Laflamme R and Negrevergne C 2002 {\it  Nature (London)} {\bf 418} 59;  Miquel C, Paz J P and Saraceno M 2002 {\it  Phys. Rev. A} {\bf 65} 062309;  Paz J P 2002 {\it  Phys. Rev. A} {\bf 65} 062311; Paz J P, Roncaglia A J and M. Saraceno 2004 Qubits in phase space: Wigner function approach to quantum error correction and the mean king problem {\it Preprint} quant-ph/0410117 

\bibitem{TH} Takami A, Hashimoto T, Horibe M and  Hayashi A 2001 {\it Phys. Rev. A} {\bf 64} 032114; Horibe M, Takami A, Hashimoto T and Hayashi A 2002 {\it  Phys. Rev. A} {\bf 65} 032105 

\bibitem{MC} Mukunda N, Chaturvedi S and Simon R 2004 {\it Phys. Lett. A} \textbf{321} 160

\bibitem {LP} Luis A and Pe\v rina J 1998 {\it J. Phys. A} {\bf 31} 1423

\bibitem{AD} Arg\"uelles A and Dittrich T 2005 Wigner function for discrete phase space: exorcising ghost images {\it Preprint} quant-ph/0504210 

\bibitem{CE} Chaturvedi S, Ercolessi E, Marmo G, Morandi G, Mukunda N and Simon R 2005 Phase-space descriptions of operators and the Wigner distribution in quantum mechanics I. A Dirac inspired view {\it Preprint} 
  
\bibitem{Dir} Dirac P A M 1945 {\it Revs. Mod. Phys.} \textbf{17} 195 and 1947 {\it The Principles of Quantum Mechanics} (Oxford, at the Clarendon Press, 3d edition)

\bibitem{Iva} Ivanovic I D 1981 {\it  J. Phys. A} {\bf 14} 3241; Wootters W K and  Fields B D 1989 {\it Ann. Phys. (N.Y.)} {\bf 191} 363; Calderbank A R, Cameron P J, Kantor W M and  Seidel J J 1997 {\it Proc. London. Math. Soc.} {\bf  75} 436; Bandyopadhyay S, Boykin P O, Roychowdhury V and Vatan F 2002 {\it  Algorirhmica} {\bf 34} 512; Lawrence J, Brukner C and Zeilinger A 2002 {\it Phys. Rev. A} {\bf 65} 032320; Chaturvedi S 2002 {\it Phys. Rev. A} {\bf 65} 044301; Pittenger A O and  Rubin M H 2004 {\it Linear Alg. Appl.} {\bf 390} 255; Klappenecker A and R\"otteler M 2004 {\it Lecture Notes in Computer Science}  {\bf 2948} 137; Parthasarathy K R 2004 On Estimating the State of a Finite Level Quantum System {\it Preprint} quant-ph/0408069 

\bibitem{Gib} Wootters W K 2004 {\it IBM J. of Research and Development} {\bf 48} 99;  Gibbons K S,  Hoffman M J and  Wootters W K 2004 {\it Phys. Rev. A} {\bf 70} 062101; Wootters W K 2004 Quantum measurements and finite geometry {\it Preprint} quant-ph/0406032 

\bibitem{SP} Saniga M, Planat M and Rosu H 2004 {\it J. Opt.  Quantum Semiclass. B} {\bf 6} L19; Wocjan P and Beth T 2005 {\it Quantum Information and Computation} {\bf 5} 93; Hayashi A, Horibe M and Hashimoto T 2005  Mean king's problem with mutually unbiased bases and orthogonal Latin squares {\it Preprint} quant-ph/0502092; Bengtsson I 2005 {\it AIP Conf. Proceedings} {\bf 750} 63; Barnum H 2002 Information-disturbance tradeoff in quantum measurement on the uniform ensemble and on the mutually unbiased bases {\it Preprint} quant-ph/0205155;  Klappenecker A and  R\"otteler M 2005  Mutually Unbiased Bases are Complex Projective 2-Designs {\it Preprint} quant-ph/0502031; Zauner G 1999 {\it Quantumdesigns: Grundz\"uge einer nichtkommutativen Designtheorie} Ph.D. thesis (Universit\"at Wien) 

\bibitem{MS} Mukunda N and Simon R 1993 {\it Ann. Phys. (N.Y.)} {\bf 228} 205

\bibitem{Pra} Mukunda N 1978  Pramana {\bf 11} 1 

\bibitem{Fir} Wootters WK  first of Refs. \cite{Woo}

\end{thebibliography}
\end{document}